# Pure iron grains are rare in the universe

Y. Kimura[1]*, K. K. Tanaka[1], T. Nozawa[2], S. Takeuchi[3], Y. Inatomi[3.4]

[1] Institute of Low Temperature Science, Hokkaido University, Sapporo 060-0819, Japan

[2] Division of Theoretical Astronomy, National Astronomical Observatory of Japan, Osawa 2-21-1, Mitaka, Tokyo 181-8588 Japan

[3] Japan Aerospace Exploration Agency, Institute of Space and Astronautical Science, 3-1-1 Yoshinodai, Chuo-ku, Sagamihara, Kanagawa, 252-5210, Japan

[4] SOKENDAI (The Graduate University for Advanced Studies), School of Physical Sciences, 3-1-1 Yoshinodai, Chuo-ku, Sagamihara, Kanagawa, 252-5210, Japan

*Correspondence to: ykimura@lowtem.hokudai.ac.jp

**Abstract**

The abundant forms in which the major elements in the universe exist have been determined from numerous astronomical observations and meteoritic analyses. Iron (Fe) is an exception, in that only depletion of gaseous Fe has been detected in the interstellar medium, suggesting that Fe is condensed into a solid, possibly the astronomically invisible metal. To determine the primary form of Fe, we replicated the formation of Fe grains in gaseous ejecta of evolved stars by means of microgravity experiments. We found that the sticking probability for formation of Fe grains is extremely small; only several atoms will stick per hundred thousand collisions, so that homogeneous nucleation of metallic Fe grains is highly ineffective, even in the Fe-rich ejecta of Type Ia supernovae. This implies that most Fe is locked up as grains of Fe compounds or as impurities accreted onto other grains in the interstellar medium.

**One Sentence Summary**

The extremely low sticking probability of iron inhibits the formation of metallic iron grains around evolved stars.

**Introduction**

Most of the atoms in the interstellar medium exist in the gas phase. Only about 1% of the total mass of the elements form tiny solid particles, called grains. Despite their low abundance, these grains are significant as building blocks of planetary systems (*1*), as substrates for the formation of molecules (*2, 3*), as energy transducers in interstellar and circumstellar environments (*4, 5*), and as key players in the efficient formation of stars (*6, 7*). The efficiency of these various contributions depends strongly on the chemical composition, size, crystal structure, and geometry of the grains that are initially formed in gaseous outflows of evolved stars and subsequently processed in interstellar environments. Therefore, to understand these characteristics of grains, it is necessary first to clarify the compositions and quantities of grains formed in stellar gas outflows.

Iron is a key element for deciphering the overall composition and amount of interstellar grains, because it is the most abundant refractory element in concurrence with magnesium and silicon in the cosmic abundance (*8*). Possible major components of Fe-rich grains include metallic iron, iron oxide, or iron sulfide (9). Fe-bearing grains are highly efficient catalysts for the molecular formation (*10*). In addition, some Fe-bearing grains have high magnetic susceptibilities, and the resulting polarized thermal emissions and magnetic dipole emissions might efficiently disturb the cosmic microwave background (*11*). Therefore, the identification of the most common form of Fe is crucial in understanding the evolution history of grains, the reprocessing of electromagnetic waves, and the chemistry in the universe.

Despite extensive astronomical observations and analyses of meteorites, insufficient amounts of Fe compounds, including iron oxides, sulfides, and carbides, have been detected to account for the expected abundance of Fe in the universe (*12-14*), suggesting that most cosmic Fe atoms exist as grains of the pure metal. Therefore, the feasibility of formation of metallic Fe grains in astronomical environments is an important subject for study. Here, to elucidate the likelihood of formation of Fe grains, we perform the reproduction experiment for the condensation of Fe grains in microgravity as a model system.

**Results**
**Advantages of the microgravity experiment**

Condensation of grains from the gas phase proceeds through the nucleation of stable small clusters and their subsequent growth. These processes are mainly controlled by two physical quantities: the sticking probability with which gas-phase atoms attach onto clusters or grains, and the surface tension of small clusters. In most theoretical models of grain formation, the sticking probability has been assumed to be 1 and the surface tension has been assumed to equal that of the corresponding bulk material. However, this optimistic assumption regarding the sticking probability may lead to an overestimation of the grain-formation rate; furthermore, the surface tension of particles with sizes of less than a few nanometers must differ markedly from the corresponding bulk value (*15, 16*). To determine the physical quantities involved in the formation of Fe grains, we performed an ideal nucleation experiment in a microgravity environment of $(6.3 \pm 0.8) \times 10^{-4}$ *G* (Fig. S1) aboard the sounding rocket *S-520-28*. An on-board *in situ* observation system composed of a nucleation chamber with a double-wavelength Mach–Zehnder-type laser interferometer and an image-recording system was adapted to conform with the size and weight limitations of the rocket (Figs. 1 and S2; *16, 17*).

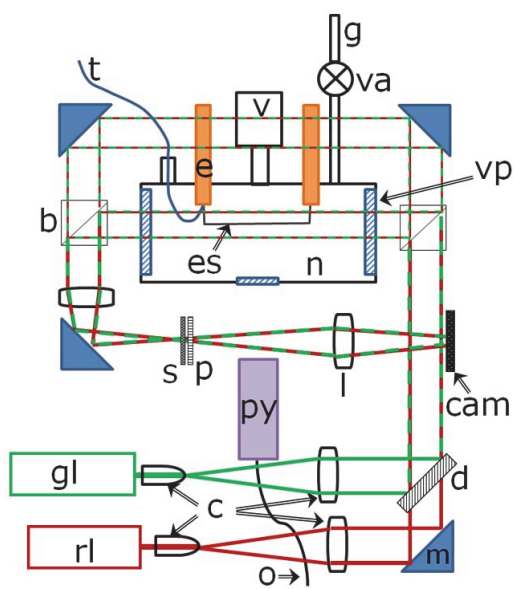

**Fig. 1. Schematic of the configuration and optical path of the double-wavelength Mach–Zehnder-type laser interferometer with a nucleation chamber**. The red and green lines show the optical paths of the red and green lasers, respectively. The resulting images of interference fringes are recorded with a CCD camera (cam). The evaporation source of Fe wire wrapped around a tungsten filament 0.3 mm in diameter and 68 mm in length is shown as the black solid line, es, in the nucleation chamber, n. The other labels are as follows: b, beam splitter; c, collimator; d, dichroic mirror; e, electrode; g, gas line; l, lens;

m, mirror; o, optical fiber; p, polarizer; s, short-pass filter; t, thermocouple; v, vacuum gauge; gl, green laser; py, pyrometer; rl, red laser; va, valve; vp, view port.

Nucleation experiments are conducted by observing the condensation of evaporated gas from a heated source in a buffer argon (Ar) gas. The Ar gas reduces the mean free path of the gas, thereby permitting a reduction in the physical scale of the experimental system. In ground-based experiments subject to the Earth's gravity, conditions for nucleation are not uniform because of the presence of thermal convection generated by the heated source. In a microgravity environment of the order of $\sim 10^{-4}$ $G$, this thermal convection is fully suppressed, conferring the following three advantages. First, the evaporated hot gas diffuses isotropically and the temperature profile around the evaporation source becomes concentric; consequently, nucleation occurs concentrically as confirmed by microgravity experiments conducted in an aircraft (Fig. S3). Secondly, the gases cool more slowly and consequently gaseous atoms can collide with each other more frequently on a longer timescale of gas cooling, providing a closer simulation of astronomical environments such as the ejecta of supernovae or outflows from asymptotic giant branch stars. The nucleation process can be approximately described in terms of the product of the timescale for the supersaturation increase ($\tau_{sat}$) and the collision frequency of iron atoms ($v$); $\tau_{sat} v$ is about $10^3$–$10^4$ for grain formation in supernovae and asymptotic giant branch stars (*18, 19*). A similar value of $\tau_{sat} v$ of $\sim 10^4$ can be achieved in microgravity experiments (see Materials and Methods), whereas $\tau_{sat} v$ is only about $10^2$ in ground-based experiments (**16, 17**). Finally, the absence of thermal convection permits the reliable determination of physical quantities by comparison of the experimental results with values from the nucleation model. In ground-based experiments, the thermal convection supplies buffer gas continuously to the nucleation sites of grains thereby complicating the interpretation of results.

**Results of the microgravity experiment**
In our microgravity experiments, the temperature and partial pressure of the Fe gas were determined simultaneously by observing shifts in the interference fringes of two lasers emitting polarized green (532 nm) and red (635 nm) lights (Fig. 2A-C and Fig. S4). When the Fe evaporation source was electrically heated on a tungsten filament, the initial interference fringes in Fig. 2A were shifted due to a decrease in the refractive index of the warmed Ar buffer gas that filled the chamber at an initial pressure of $4.0 \times 10^4$ Pa (e.g., Fig. 2B). When the source temperature was increased to $2226 \pm 22$ K, the interference fringes in the upper right corner of Fig. 2C disappeared. Nucleation of Fe grains can be detected from this disappearance of interference fringes, which results from scattering of the incident laser light by the newly formed Fe grains. We define the position where this occurs as the nucleation front (indicated by the dotted line in Fig. 2C), at which Fe grains nucleate and grow immediately. The derived temperature and partial pressure of Fe gas at the nucleation front were $907 \pm 20$ K and $(2.23 \pm 0.27) \times 10^3$ Pa, respectively; the uncertainty in these values arose from inaccuracies in measurements of the shifts of the interference fringes (see Materials and Methods). The temperature and partial pressure of Fe just before the nucleation decreased smoothly from the evaporation source to the nucleation front, whereas the number density of Fe was constant (Fig. 2D).

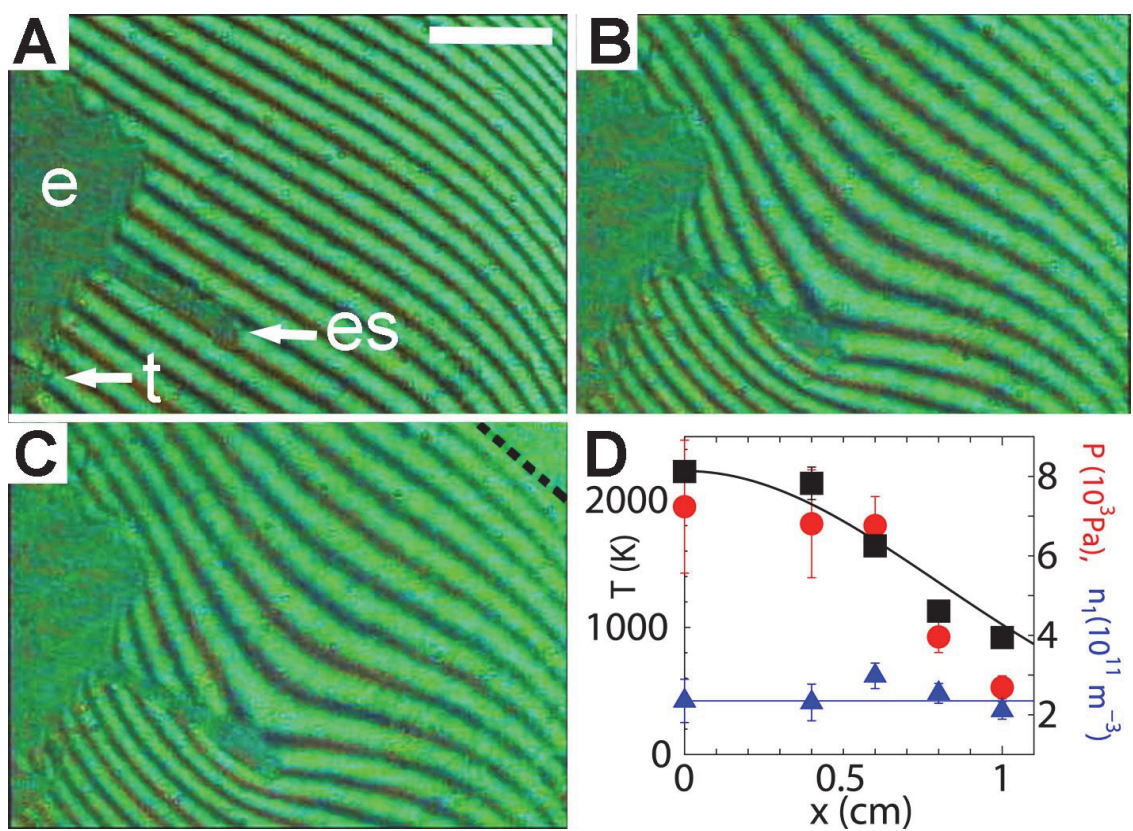

**Fig. 2. Photographs of interference images and the temperature and partial pressure during the experiment on Fe nucleation under microgravity.**
Color images of the interference fringes (see Fig. S4 for monochromatic images) at representative times for the experiment with an initial pressure of Ar buffer gas of $4 \times 10^4$ Pa: (**A**) before heating of the evaporation source, (**B**) 0.4 s before the nucleation of Fe grains and (**C**) at the time of nucleation. In (**C**), the dotted line in the upper right corner indicates the nucleation front of the Fe grains above which the disappearance of interference fringes is due to scattering of light by abundantly formed Fe grains. The panel (**D**) plots profiles of the temperature (squares), partial pressure (circles) and number density (triangles) of the Fe gas from the evaporation source to the nucleation front in panel **B**. The error for the *x*-axis is within the symbols. The solid black and blue curves are, respectively, the temperature profile and the initial number density of Fe atoms used in the calculation. The temperature was expressed by Eq. (10) with a time $t = x^2 D^{-1}$, where $x$ is the distance from the evaporation source and $D$ is the diffusion coefficient of Fe atoms. The labels are as follows; e, electrode; t, thermocouple; es, evaporation source. The scale bar in **A** corresponds to 3 mm.

To examine the effects of Ar buffer gas, we performed an additional experiment at a reduced Ar pressure of $2.0 \times 10^4$ Pa. The temperature and partial pressure of Fe at the nucleation front were then $958 \pm 37$ K and $(1.44 \pm 0.29) \times 10^3$ Pa, respectively, with an elevated source temperature of 2188 K.

Figure 3A shows the temperatures and partial pressures of Fe gas at the nucleation front determined from the experiments. In the case of an initial Ar gas pressure of $4.0 \times 10^4$ Pa, the nucleation temperature of Fe grains (907 K) was significantly below the solid–vapor equilibrium temperature (2116 K) corresponding to the partial pressure of Fe $[(2.23 \pm 0.27) \times 10^3]$ at the nucleation front (Fig. 3B). The degree of supercooling is therefore as much as 1209 K. Because the equilibrium vapor pressure between gaseous and solid Fe at 907 K is $3.1 \times 10^{-11}$ Pa, an extremely high supersaturation ratio of $7.2 \times 10^{13}$ is realized at the nucleation front. For the experiment at an initial Ar gas pressure of $2.0 \times 10^4$ Pa, the resulting supersaturation ratio was $2.4 \times 10^{12}$, based on the equilibrium vapor pressure between solid and vapor at the nucleation temperature of 958 K.

**Determination of the physical properties**

By applying nucleation theory, we determined the sticking probability $\alpha$ for Fe grain formation to explain the nucleation temperature obtained in the experiments. We used a semi-phenomenological (SP) model in which the formation energy of a cluster in the classical nucleation theory (CNT) is corrected for the binding energy of dimers (see Materials and Methods). In our calculations for grain formation, we defined the nucleation temperature as the point at which half the initial gas-phase Fe atoms have been consumed, because nucleation was detected experimentally from the scattering of laser light. Consequently, a significant fraction of the Fe gas had to be locked up in Fe grains. Figure 4A shows a result of a simulation with $\alpha = 1.8 \times 10^{-5}$ and the surface tension ($\sigma$) of bulk Fe (2.48 N m$^{-1}$) (*20*). As the gas cools, the nucleation rate increases dramatically because of the increase in the supersaturation ratio. Consequently, the number density of the remaining Fe gas decreases and the nucleation rate decreases rapidly. The nucleation rate reaches a maximum at 1012 K, and the nucleation temperature is calculated to be 907 K. Therefore, this model with $\alpha = 1.8 \times 10^{-5}$ reproduces the experimental results closely. By taking into account inaccuracies in temperature, the sticking probability of Fe at ~900 K is found to be (1.4–2.0) $\times 10^{-5}$. The resulting sticking probability depends on the definition of the nucleation temperature; for example, $\alpha = 1.2 \times 10^{-5}$ when the nucleation temperature is defined as the point at which 10% of the initial gas atoms are consumed or $\alpha = 1.2 \times 10^{-4}$ at 90% consumption. Note that the grain-formation model with $\alpha = 1$ predicts a nucleation temperature of about 1750 K (Fig. 4A) and cannot therefore explain the experimental results.

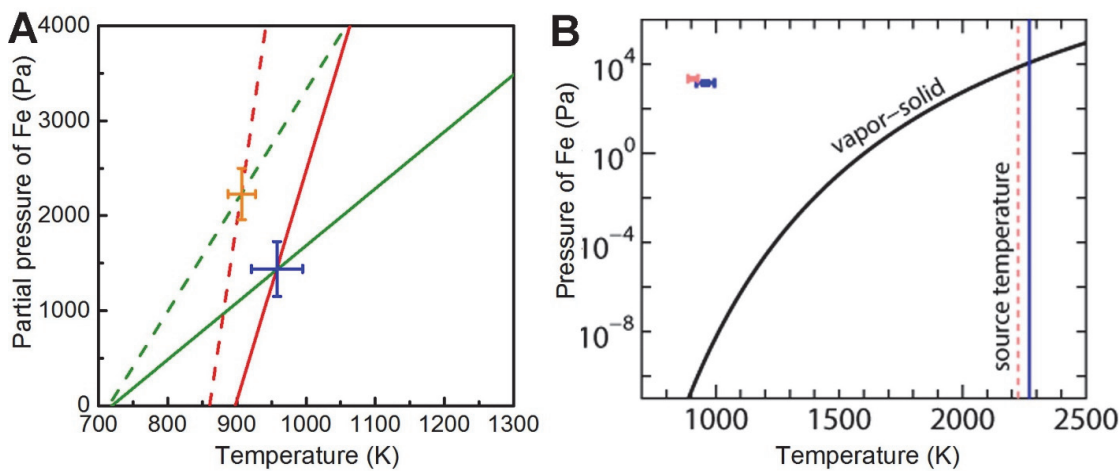

**Fig. 3. Temperatures and partial pressures of Fe gas at the nucleation front, obtained from the microgravity experiments.** (**A**) The blue and orange points with error bars indicate the temperatures and partial pressures of Fe gas just before the nucleation for the experiments at initial Ar gas pressures of $2.0 \times 10^4$ Pa and at $4.0 \times 10^4$ Pa, respectively. The green and red lines show the relationship between the temperature and the partial pressure of Fe gas to explain the shifts in the interference fringes for the green and red laser beams; the solid and dashed lines are the results for Ar gas at $2.0 \times 10^4$ Pa and $4.0 \times 10^4$ Pa, respectively. (**B**) The equilibrium vapor pressure of Fe between the vapor

and solid is shown by the solid curve. The two square symbols are the same as the points in Panel A. The temperatures measured at the evaporation source are shown by the blue and orange vertical lines for experiments in Ar gas at $2.0 \times 10^4$ Pa and $4.0 \times 10^4$ Pa, respectively. The large gap between the two square symbols and the solid curve shows the presence of a very large supersaturation at nucleation.

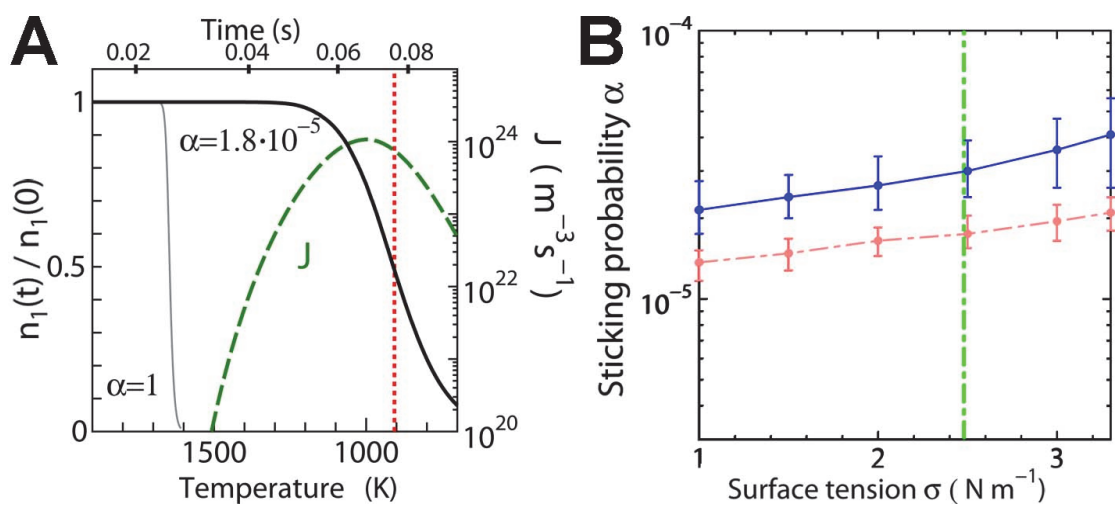

**Fig. 4. Estimation of the sticking probability by simulations to explain the results of the experiments**. (**A**) Result of calculations for the formation of Fe grains for a sticking probability of $\alpha = 1.8 \times 10^{-5}$ and a surface tension of $\sigma = 2.48$ N m$^{-1}$, obtained by applying the SP nucleation model. The dashed and solid curves show the time variation in the nucleation rate $J$ and the number density of gas-phase Fe atoms $n_1(t)$, respectively. The vertical dotted line shows the nucleation temperature derived from microgravity experiments. The thin gray curve shows the number density of Fe atoms for $\alpha = 1$ and $\sigma = 2.48$ N m$^{-1}$, for which nucleation occurs at a much higher temperature (1700 K) than the experimental result. (**B**) Plot of the sticking probability $\alpha$ against the surface tension $\sigma$ estimated from a comparison of the results of experiments and the simulations. The blue and orange points plot results for experiments in Ar gas at $2.0 \times 10^4$ Pa and $4.0 \times 10^4$ Pa, respectively. The vertical dot–dashed green line shows the bulk surface tension of molten Fe (*20*).

We also examined the dependence of the results on the surface tension. Because the surface tension $\sigma$ of nanometer-scale Fe particles is likely to differ several tens of percent from that of bulk Fe (2.48 N m$^{-1}$), we considered a wider range of values σ = 1.0–3.3 N m$^{-1}$.   As shown in **F**ig. 4B, the variability of the sticking probability is small in the plausible range of surface tensions of Fe.   As mentioned above, the supersaturation ratio is remarkably high for nucleation of Fe grains in our experiments.   As a result, the size of critical nuclei, which is the minimum number of atoms present in a small cluster that can grow continuously and with thermodynamic stability, should be small and is actually the dimer, as can be evaluated by means of Eq. 14.   In fact, the formation of dimers from isolated atoms is the largest barrier to grain formation, because the forming dimer dissociates into the gas phase using the excess energy from the bonding.   Therefore, the formation energy of a cluster is mainly determined by the binding energy of the dimer rather than by the surface tension, leading us to conclude that the binding energy of the dimer, as well as the sticking probability, is crucial for homogeneous nucleation.

**Discussion**

The extremely low sticking probability of Fe suggests that homogeneous nucleation of metallic Fe grains is highly inefficient.   A similarly small sticking probability (α = ~10$^{-5}$) has been reported for the formation of metallic zinc grains in a microgravity nucleation experiment (*21*).   We have also measured a small sticking probability (α = 3.4 × 10$^{-5}$ ± 1.2 × 10$^{-4}$) for metallic nickel grains in a microgravity experiment performed in an aircraft.   In contrast, the sticking probability for the formation of grains including Fe in ground-based laboratory experiments is significantly larger (α = ~10$^{-2}$ to 1) (*16, 22*).   How then does gravity affect the sticking probability? One possibility is localized enhancement of gas density as a result of thermal convection; a higher density leads to a higher collision frequency of atoms, which might cause overestimation of the sticking probability.   Another possibility is a heterogeneous effect; for example, small amounts of residual oxygen and water gas might be continuously supplied to the nucleation region by thermal convection.   If seed nuclei of Fe oxides form rather than metallic Fe, atomic Fe might be able to stick to Fe oxides more efficiently than to metallic Fe.

The primary sources of cosmic Fe are considered to be Type Ia supernovae, which are driven by thermonuclear fusion of white dwarfs (*23*).   Iron atoms produced around the centers of Type Ia supernovae are injected into the interstellar medium as cooling Fe-rich ejecta, in which Fe gas is believed to condense as pure Fe grains (*24*). However, despite the production of large quantities of Fe atoms, the formation of Fe grains has not yet been observed in Type Ia supernovae (*25*). A theoretical model of grain formation shows that small amounts of Fe grains with a radius of 10 nm could form in Type Ia supernovae if the sticking probability is assumed to be 1 (*26*). Our results, however, show the sticking probability for formation of Fe grains is extremely low, suggesting that the homogeneous nucleation of Fe grains is much more difficult than previously expected. This explains why no metallic Fe grains have been observed in Type Ia supernovae. In addition, the ejecta of Type Ia supernovae are subjected to strong radiation fields for up to several hundred days after the explosion; the energetic photons and electrons could destroy small clusters, causing additional suppression of grain condensation.

If this is so, where in the universe is the Fe? Core-collapse supernovae arising from massive stars also produce Fe atoms and disperse them into the interstellar medium (*27*). It has been argued that an abundant mass of metallic Fe grains formed in the ejecta of SN 1987A is required to explain the far-infrared flux observed by the *Herschel Space Observatory* (*28*). However, even if core-collapse supernovae synthesize Fe grains with high efficiency, we cannot conclude that cosmic Fe exists predominantly in the form of pure Fe grains, because core-collapse supernovae are not the dominant source of Fe in the present universe (*27*). In addition, given the very low sticking probability determined from our experiments, Fe may not condense as the pure metal but, instead, it may condense onto other species of grains through heterogeneous nucleation. Many studies suggest that grains can grow through accretion of gas-phase atoms in interstellar environments such as dense molecular clouds. In this case, it would be unnatural for Fe gas to accrete predominantly onto Fe grains rather than onto existing silicate or carbonaceous grains (*29, 30*). Therefore, most of the Fe might be captured as impurities and/or components of other grains through physical and chemical processes in the interstellar medium.

**Materials and Methods**

**Determination of the temperature and partial pressure of Fe**

We determine the partial pressure of evaporated Fe gas by detecting small changes in the refractive index by individually observing shifts in the interference fringes at two wavelengths. The refractive indices of Ar, $N_{Ar(T,P)}$, and Fe, $N_{Fe(T,P)}$, can be expressed as a function of the temperature $T$ (K) and pressure $P_{gas}$ (Pa), as follows:

$$N_{gas(T,P)} - 1 = \frac{[N_{gas(273.15,P_0)} - 1]}{1 + a\Delta T} \frac{P_{gas}}{P_0}, \quad (1)$$

where $a$ is the coefficient of volume expansion (0.003663 K$^{-1}$ for Ar and 0.003661 K$^{-1}$ for Fe in this experiment), $\Delta T =$ T − 273.15 K, and pressure $P_0$ = 101325 Pa. The values of $N_{Ar}$ − 1 are (2.790 ± 0.017) × 10$^{-5}$ at 632.8 nm and (2.813 ± 0.017) × 10$^{-5}$ at 532 nm, respectively, at 1.0 × 10$^4$ Pa and 293.15 K (*31*). The values of $N_{Fe}$ − 1 are 3.837 × 10$^{-4}$ and 1.163 × 10$^{-4}$ at 532 nm and 633 nm, respectively, at 2.0 × 10$^4$ Pa and 293 K (*32*).

The product of the shift in the fringes, $\Delta d$, and the wavelength of the laser, $\lambda$, is proportional to the change in the optical path length $L$, defined as $L = Nl$, where $l$ is the physical length (taken to be the length of the tungsten filament 68 mm in this experiment). The shifts in the positions of the interference fringes for the green ($\Delta d_G$) and red lasers ($\Delta d_R$) after heating are given by the following equations:

$$\Delta d_G = [N_{G,Ar(Ti,Pi)} - N_{G,Ar(T,P-P_{Fe})} - N_{G,Fe(T,P_{Fe})} + 1]\frac{l}{\lambda_G} \quad (2)$$

and

$$\Delta d_R = [N_{R,Ar(Ti,Pi)} - N_{R,Ar(T,P-P_{Fe})} - N_{R,Fe(T,P_{Fe})} + 1]\frac{l}{\lambda_R}, \quad (3)$$

respectively, where $T_i$ and $P_i$ are the initial temperature and pressure of Ar before the source temperature was elevated, and the subscripts $G$ and $R$ indicate quantities for the green and red lasers, respectively. Because the total pressure in the chamber $P$ was monitored by using a pressure gauge, and $\Delta d_G$ and $\Delta d_R$ can be observed in the images, it was possible to determine the gas temperature ($T$) and the partial pressure of Fe ($P_{Fe}$) simultaneously by using Eqs. (1–3). The amounts of fringe shift at the nucleation front just before the nucleation were $\Delta d_G$ = 6.5 and $\Delta d_R$ = 6.4, respectively, for the experiment at an initial Ar gas pressure of 4 ×10$^4$ Pa.

**Model and Numerical Simulation**

To interpret the results of our experiments, we performed numerical simulations of the non-equilibrium condensation of Fe on the basis of nucleation theory (*21, 33*), by applying a semi-phenomenological (SP) nucleation model which is one of the most successful and useful models (*34, 35*). The SP model modifies the CNT by adding extra terms for the formation energy of a cluster, obtained by using the second virial coefficient of the vapor. This model succeeded in achieving good agreement with the nucleation rates derived from experimental data or from MD simulations for various materials (*34–37*). In homogeneous nucleation theory, the nucleation rate $J$ (i.e., the rate at which the stable nuclei are formed per volume) is expressed as follows:

$$J = \left\{ \sum_{i=1}^{\infty} \frac{1}{R^+(i) n_e(i)} \right\}^{-1}, \quad (4)$$

where $R^+(i)$ is the accretion rate from an $i$-mer (a cluster containing $i$ atoms) to an $(i+1)$-mer. The equilibrium number density of $i$-mers $n_e(i)$ is given by

$$n_e(i) = \frac{P_{Fe}}{kT} \exp\left(-\frac{\Delta G_i}{kT}\right), \quad (5)$$

where $P_{Fe}$ is the partial pressure of Fe gas, $k$ is the Boltzmann constant, and $\Delta G_i$ is the free energy associated with the formation of a cluster of size $i$ from the gas phase. The accretion rate $R^+(i)$ is given by

$$R^+(i) = \alpha n_1 v_{th} \left(4\pi r_1^2 i^{\frac{2}{3}}\right), \quad (6)$$

where $n_1$ is the number density of Fe atoms, $v_{th}$ is the thermal velocity of gas given by $\sqrt{kT/(2\pi m)}$, and $\alpha$ is the sticking probability. The radius of an atom $r_1$ is defined as $(3m/4\pi\rho_m)^{1/3}$, where $m$ is the mass of an atom and $\rho_m$ is the bulk density.

Because of the exponential dependence of the free energy on $J$, evaluation of the free energy $\Delta G_i$ is important. In this study, we used values of $\Delta G_i$ given by an SP model. In the SP model, $\Delta G_i$ is expressed as

$$\Delta G_i = -(i-1)kT \ln S + \sigma A_1 \left(i^{\frac{2}{3}} - 1\right) + A_2 \left(i^{\frac{1}{3}} - 1\right), \quad (7)$$

where $S (= P_{Fe}/P_{Fe,e})$ is the supersaturation ratio, $\sigma$ is the surface tension, $A_1$ is the surface area of a monomer and $A_2$ is a correction term determined from the second virial coefficient. $P_{Fe,e}$ is the equilibrium vapor pressure of bulk Fe materials at a given temperature (Fig. 3B).

Although we have no experimental data on the second virial coefficient of refractory metals such as Fe, we can evaluate $A_2$ by using a relationship between the second virial coefficient and the chemical potential of dimer. According to Tanaka *et al.* (*33*), $A_2$ is expressed as

$$A_2 = \left[-\left(2^{2/3}-1\right)\sigma A_1 - E + \mu_V + \mu_R + \mu_T\right]\left(2^{1/3}-1\right)^{-1}, \qquad (8)$$

where $E$ is the binding energy of a dimer and $\hbar$ is Plank's constant (*32*). In Eq. (8), $\mu_V$, $\mu_R$ and $\mu_T$ are the chemical potentials arising from vibrational, rotational and translational motions of the dimers, as given by

$$\mu_V = kT \ln\left\{1 - \exp\left(-\frac{\hbar\omega}{kT}\right)\right\} + \frac{\hbar\omega}{2}, \quad \mu_R = -kT \ln\left(\frac{mR_e^2 kT}{2\hbar^2}\right), \quad \mu_T = kT \ln\left\{\frac{P_{Fe,e}}{kT}\left(\frac{mkT}{4\pi\hbar^2}\right)^{-3/2}\right\}, \quad (9)$$

where $\omega$ is the vibrational frequency of a dimer and $R_e$ is the equilibrium distance between nuclei. Consequently, we can evaluate $A_2$ as a function of $T$ for given $\sigma$, $E$, $\omega$, and $R_e$. In our calculation, we adopted the values $E/k = 8600$ K, $\omega = 8.9 \times 10^{12}$ s$^{-1}$, and $R_e = 2.4 \times 10^{-10}$ m (*38*). In the SP model, the formation energy of a cluster is determined by the binding energy and chemical potentials arising from vibrational, rotational, and translational motions of dimers, and the bulk surface tension. The formation energy for small clusters is primarily determined by the binding energy of the dimer rather than by the bulk surface tension, because the value of $E$ is markedly dependent on the material. Note that the SP model gives the exact values of the free energy of monomer and dimer ($\Delta G_1$ and $\Delta G_2$), because $\Delta G_1 = 0$ is satisfied and $\Delta G_2$ corresponds to the chemical potential of the dimer. There are large deviations from the values evaluated by CNT for clusters of fewer than about ten atoms. Consequently, the correction term in $\Delta G_i$ is crucial for small clusters.

In the calculations, we consider a gaseous system containing Fe gas that cools with a characteristic time $\tau_T$. In this case, the temperature $T$ of the gas as a function of time $t$ is given by

$$T(t) = T_0 \exp(-t/\tau_T), \qquad (10)$$

where $T_0$ is the initial temperature, which corresponded to the temperature of the heated source in our calculations (see below). As the gas temperature decreases, nucleation and grain growth proceed, consuming gas-phase Fe atoms. The number density of Fe gas $n_1(t)$ is given by

$$n_1(t) = n_1(0) - \int_0^t J(t') \left(\frac{r(t,t')}{r_1}\right)^3 dt', \quad (11)$$

where $J(t')$ is the nucleation rate at time $t'$, and $r(t,t')$ is the radius of grains nucleated at $t'$ and measured at $t$. The growth equation of a grain radius $r(t,t')$ is expressed in the form

$$\frac{\partial r(t,t')}{\partial t} = \alpha \frac{4\pi}{3} r_1^3 n_1(t) v_{th}. \quad (12)$$

The radius of the critical nuclei $r(t',t')$ is

$$r(t,t') = i_*^{1/3} r_1, \quad (13)$$

where $i_*$ is the number of atoms in the critical cluster, which is the smallest thermodynamically stable clusters and determined by $dn_e(i)/di = 0$; i.e.,

$$i_* = \left(\frac{\sigma A_1 + \sqrt{(\sigma A_1)^2 + 3A_2 kT \ln S}}{3kT \ln S}\right)^3. \quad (14)$$

By using Eqs. (4–14), which describe the nucleation rate and the non-equilibrium condensation, we simulate the condensation process of Fe by treating the sticking probability $\alpha$ and the surface tension $\sigma$ as parameters.

In the calculations, we adopted initial partial pressures of Fe of $P_{Fe\,(t=0)}$ (= $n_1(0)kT$) = 1440 and 2230 Pa and initial temperatures of $T_0$ = 2188 and 2226 K for the experiments in Ar gas at initial pressures of $2.0 \times 10^4$ and $4.0 \times 10^4$ Pa, respectively. The timescale for cooling $\tau_T$ was taken as the time required for the Fe gas to arrive at the nucleation site by diffusion from the evaporation source: $\tau_T \approx X^2 D^{-1}$, where $X$ is the distance from the evaporation source to the nucleation site ($X$ = $1.29 \times 10^{-2}$ and $1.14 \times 10^{-2}$ m), and $D = v_{mean}\lambda/3$ is the diffusion coefficient ($D = 3.07 \times 10^{-3}$ and $1.56 \times 10^{-3}$ m$^2$ s$^{-1}$). Here, $v_{mean} = (8kT/\pi m)^{1/2}$ is the mean velocity of the gas ($v_{mean}$ = 783 or 771 m s$^{-1}$), and $\lambda = (\sqrt{2}ns)^{-1}$ is the mean free path of a gas molecule ($\lambda = 1.18 \times 10^{-5}$ or $6.06 \times 10^{-6}$ m). The mean cross-section of an Fe atom $s$ is $4.988 \times 10^{-20}$ m$^2$ and the number density of Ar gas $n$ is $1.20 \times 10^{24}$ m$^{-3}$ ($2.34 \times 10^{24}$ m$^{-3}$) for the experiments in Ar gas at an initial pressure of $2.0 \times 10^4$ Pa ($4.0 \times 10^4$ Pa), where we used the total gas pressures of 26550 Pa (50683 Pa), measured in the experiments. The obtained timescale for cooling is $\tau_T = 5.42 \times 10^{-2}$ and $8.33 \times 10^{-2}$ s for the experiments in Ar gas at initial pressures of $2.0 \times 10^4$ and $4.0 \times 10^4$ Pa, respectively. In these estimations, we adopted the mean temperatures (1614 or 1567 K) between the evaporation source and the nucleation site. There is an uncertainty in the value of $\tau_T$ within a factor of 2, because $\tau_T$ is proportional to $T^{-3/2}$, leading to small deviations from the evaluated value of $\alpha$ to explain the experiments (within a factor of 2).

It is known that the nucleation process can be approximately described in terms of the product of the timescale for the supersaturation increase $\tau_{sat}$ and the collision frequency of iron atoms $\nu$ (*18*). The collision frequency of iron atoms $\nu$ is $2.52 \times 10^6$ or $3.96 \times 10^6$ s$^{-1}$ for experiments in Ar gas at initial pressures of $2.0 \times 10^4$ and $4.0 \times 10^4$ Pa, respectively. Because $\tau_{sat}$ is evaluated from $\tau_T kT/H$ with the latent heat ($H/k = 4.5 \times 10^4$ K), the corresponding values of $\tau_{sat} \nu$ are calculated to be $4.9 \times 10^3$ and $1.1 \times 10^4$, which are similar to those in the grain formation environments of supernovae and asymptotic giant branch stars (*19*), whereas $\tau_{sat} \nu$ was determined to be ~$10^2$ in ground based experiments, due to the smaller value of $\tau_T$ (~$10^{-4}$ s) (*16*, *17*).

*Science Advances*, 18 January 2017:   3 (2017) e1601992
DOI: 10.1126/sciadv.1601992

*Science Advances*, 18 January 2017: 3 (2017) e1601992
DOI: 10.1126/sciadv.160199233. K. K. Tanaka, H. Tanaka, K. Nakazawa, Non-equilibrium condensation in a primordial solar nebula: Formation of refractory metal nuggets. *Icarus* **160**, 197−207 (2002).
34. A. Dillmann, G. E. A. Meier, A refined droplet approach to the problem of homogeneous nucleation from the vapor phase. *J. Chem. Phys.* **94**, 3872−3884 (1991).
35. A. Laaksonen, I. J. Ford, M. Kulmala, Revised parametrization of the Dillmann–Meier theory of homogeneous nucleation. *Phys. Rev. E: Stat. Phys., Plasmas, Fluids, Relat. Interdiscip. Top.* **49**, 5517−5524 (1994).
36. K. K. Tanaka, H. Tanaka, T. Yamamoto, K. Kawamura, Molecular dynamics simulations of nucleation from vapor to solid composed of Lennard–Jones molecules. *J. Chem. Phys.* **134**, 204313(1-13) (2011).
37. R. Angelil, J. Diemand, K. K. Tanaka, H. Tanaka, Homogeneous SPC/E water nucleation in large molecular dynamics simulations. *J. Chem. Phys.* **143**, 064507(1-10) (2015).
38. I. Shim, K. A. Gingerich, *Ab initio* HF-CL calculations of the electronic "band structure" in the $Fe_2$ molecule. *J. Chem. Phys.* **77**, 2490–2497 (1982).
**Acknowledgments**
**General**: We thank N. Ishii and all project members of the sounding rocket *S-520-28* experiment of JAXA.  **Funding:** This work was supported by a Grant-in-Aid for Scientific Research(S) from KAKENHI (15H05731), by a Grant-in-Aid for Young Scientists (A) from KAKENHI (24684033) of JSPS, by a Grant-in-Aid for Scientific Research(C) from KAKENHI (2640023, 15K05015), by a Grant-in-Aid for Scientific Research on Innovative Areas (16H00927), and by the Steering Committee for Space Biology and Microgravity Science of Institute of Space and Astronautical Science (ISAS), Japan Aerospace Exploration Agency (JAXA).  **Author contributions:** Y.K. designed the project, performed experiments and analyzed data.  K.K.T. performed the calculations.  Y.K., K.K.T., T.N. interpreted the data and co-wrote the paper.  Y.K., S.T., Y.I. prepared the experimental system.  S.T., Y.I. prepared the interface between experimental system and the sounding rocket.  **Competing interests:** The authors declare that they have no competing interests.  **Data and materials availability:** All data needed to evaluate the conclusions in the paper are present in the paper and/or the Supplementary Materials.  Additional data related to this paper may be requested from the authors.

**Supplementary Materials**

**Supplementary Text**
**Experimental System and Procedure**

   The nucleation chamber (n in Fig. S2A) was a 150-mm-long stainless-steel cylinder with an internal diameter of 65 mm.   It was equipped with two viewports for recording the paths of interference fringes of the two lasers, and one viewport for measuring the temperature of the evaporation source using a pyrometer (ISQ5-LO, Hazama Sokki Co. Ltd., Yokohama). Two ports were also equipped with a 0.1-mm-diameter chromel–alumel thermocouple [a combination of WF-1/8"PT -0.8-2-T -TK-1000 mm/150 mm (Tecsam Co. Ltd., Hsinchu) and KMT-100-100-050 (ANBE SMT Co., Yokohama)] for measuring the temperature at the end of the evaporation source as an alternative method for determining the temperature of the evaporation source.   Two electrodes (PF-SM6-3KV-10A, Kawaso Texcel Co., Osaka) were used for heating the evaporation source.   The chamber also had a specially coordinated high-resolution pressure gauge (HAV-60KP-V; Sensez Co., Tokyo).   The accuracy of pressure measurement was ±90 Pa, which corresponds to an accuracy of ±1.5 K at the nucleation temperature (~900 K) given by our experimental results.   A quarter-inch stainless steel tube (g in Fig. S2A) was connected to a vacuum system through a valve (6LVV-DPBW4-P1; Swagelok Co., Manchester; va in Fig. S2B) for evacuation and the introduction of gas.   The air in the chamber was evacuated by a combination of a turbo-molecular pump (TG50F, 50 L/s; Osaka Vacuum, Ltd.) and a scroll-type dry vacuum pump (DIS-90; ULVAC Kiko Inc., Saito City).   After a sufficient vacuum was attained (pressure $10^{-5}$ Pa), pure Ar gas (>99.9999% purity) was injected into the chamber. The Ar buffer gas was essential for measuring the partial pressure and temperature of evaporated Fe gas based on the shifts of interference fringes of optical lasers at two different wavelengths (see below).   In addition, the inert Ar gas decreased the mean free path and shortened the cooling timescale of the evaporated Fe gas, which allowed us to perform the nucleation experiments within the short duration of the microgravity environment and within the limited space available in the rocket.   To observe the effects of the Ar gas pressure on the results, the two nucleation chambers with identical configurations were installed in the rocket (Fig. S2B) and each was filled with Ar gas at a different pressure ($2 \times 10^4$ Pa and $4 \times 10^4$ Pa).

   The Mach–Zehnder-type interferometer (Fig. S2A) had two lasers: a polarized green laser with a wavelength $\lambda_G$ = 532 nm (compact green laser module, 10 mW

BEAM MATE HK-5616; Shimadzu Corp., Kyoto) and a red laser with a wavelength $\lambda_R$ = 635 nm (4.5 mW continuous wave circular beam laser diode module; Edmund Optics Inc., Barrington, NJ).  The evaporation source, a Fe wire (0.1 mm$\phi$ × 100 mm) wrapped around a tungsten filament (0.3 mm$\phi$ × 68 mm), was carefully aligned parallel to the optical path (<4 × 10$^{-4}$ rad) and was made as long as possible to obtain a high column density of evaporated Fe gas.  This enabled us to detect tiny changes in the refractive index of as little as 1.0  × 10$^{-6}$, which corresponds to a change in temperature from 298 K to 301 K for Ar gas at 2 × 10$^4$ Pa.  Because the vapor pressure of metallic tungsten is much lower (10$^{-5}$ Pa at 2508 K) than that of Fe (9.7 × 10$^4$ Pa at 2508 K), the partial pressure of evaporated metallic tungsten could be neglected during our experiments.

The interference fringes of the green and red lasers were captured by a recording system (Board Camera MS-88HCS without a low-pass filter; Moswell Co., Ltd., Yokohama) and were downloaded to the ground by telemetry at a rate of 10 frames per second during the experiment.  The spatial resolution was ~45 μm.

**Launch of sounding rocket S-520-28**

Sounding rocket *S-520-28* of the Japan Aerospace Exploration Agency (JAXA) was launched at 4 pm JST on December 17th, 2012, and reached an altitude of 312 kilometers 283 seconds after liftoff.  The mean gravitational acceleration during the parabolic flight, measured by a triaxial analogue accelerometer module (Model 2470-002; Silicon Designs Inc., Kirkland, WA) placed on a base plate of the nucleation chamber was (6.3 ± 0.8) × 10$^{-4}$ G (Fig. S1), which is two orders of magnitude lower than that in microgravity experiments performed aboard aircraft. The duration of the microgravity was as much as ~445 seconds in this sounding rocket, much longer than the ~20 seconds achievable in aircraft.  Such relatively long and better microgravity condition allowed us to perform two successive nucleation experiments at different Ar gas pressures (after waiting for residual convection to settle down) in a single flight. The rotational frequency of the rocket was kept as low as –0.018 Hz.  Electrical heating of the evaporation sources in Ar gas at an initial pressure of 2.0 × 10$^4$ Pa (4.0 × 10$^4$ Pa) was performed for 65 seconds in total by constant elevation of the applied voltage (~0.13 V s$^{-1}$) between 180 and 240 seconds (255 and 315 seconds) after the launch, and at constant voltage at 8.6 V for the subsequent 5 seconds.

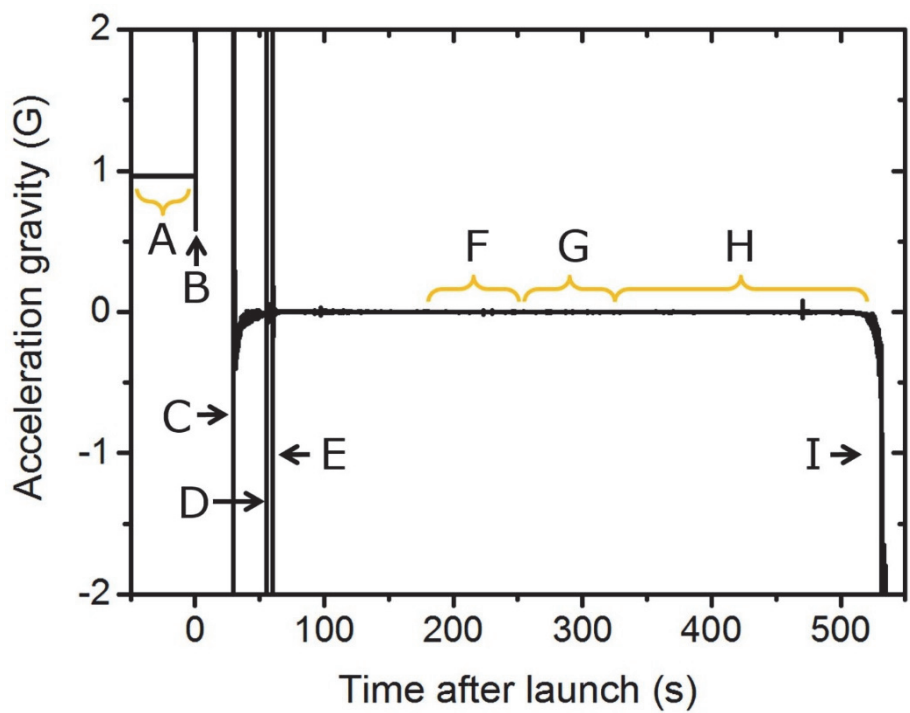

**Fig. S1. Time evolution of the acceleration gravity in the sounding rocket during the microgravity experiment.** Labels A to I indicate the following events: A, before launch; B, launch; C, termination of ignition; D, despinning by spreading of yo-yo; E, opening of the nose cone for release of heat; F, experiment in Ar gas at $2.0 \times 10^4$ Pa; G, experiment in Ar gas at $4.0 \times 10^4$ Pa; H, transfer of the recorded images to the ground; and I, termination of parabolic flight. The gravitational acceleration was measured by a triaxial analogue accelerometer module positioned on the base plate of the experimental system. It was same order in three axes during parabolic flight.

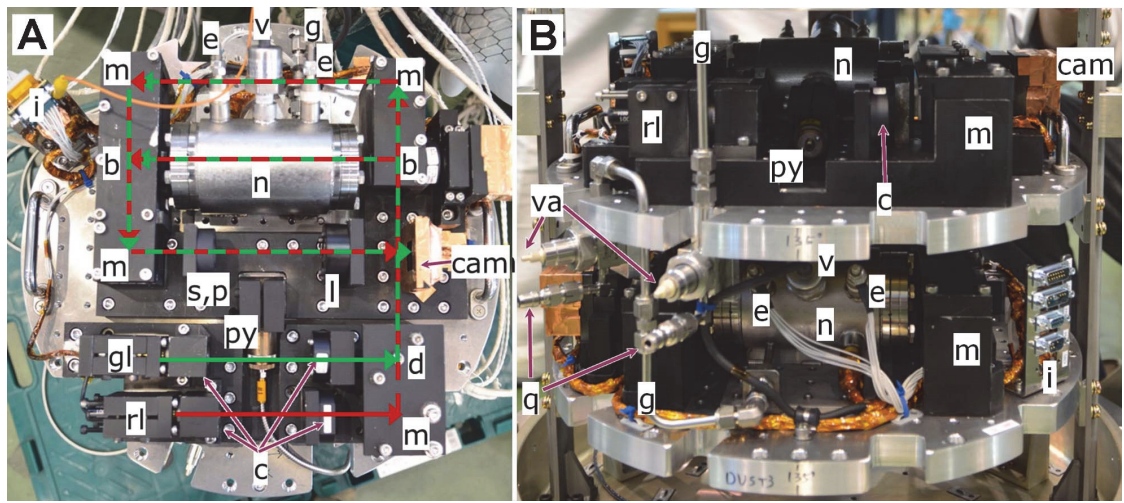

**Fig. S2. Photographs of the experimental systems.** **(A)** Top view of the double-wavelength Mach–Zehnder-type laser interferometer with the nucleation chamber.   Red and green arrows show the optical paths of red and green lasers, respectively.   **(B)** Side view of the experimental systems.   View directions of the upper and bottom systems are from bottom and top of A, respectively, i.e., each system is 180º opposite in phase.   Labels indicate: b, beam splitter; c, collimator; d, dichroic mirror; e, electrode; g, gas line; i, interface connecters; l, lens; m, mirror; n, nucleation chamber; p, polarizer; q, Quick-Connect; s, short-pass filter; v, vacuum gauge; cam, CCD camera; gl, green laser; rl, red laser; py, pyrometer and va, valve.   Evacuation of the air and subsequent injection of Ar gas into the chamber were performed on the ground after the experimental system had been installed inside the rocket.

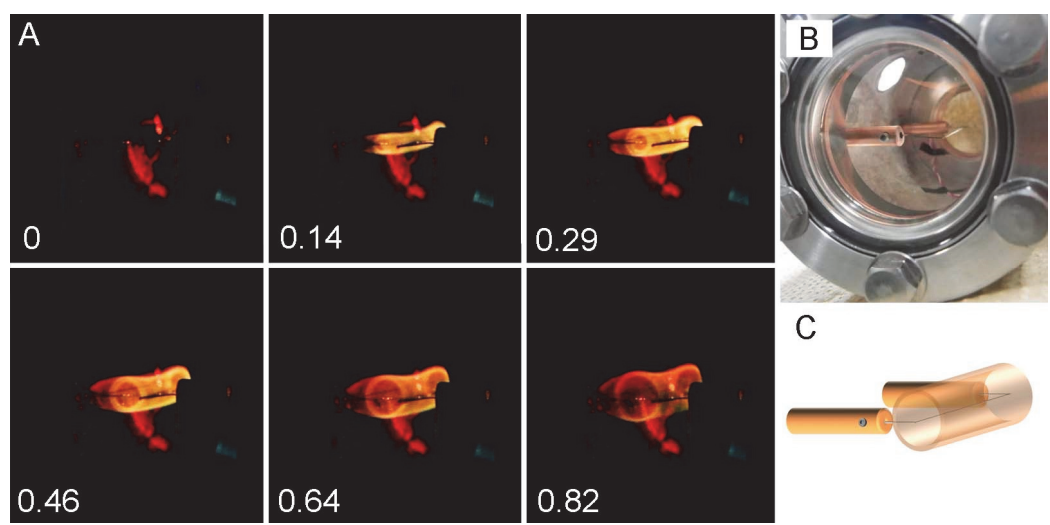

**Fig. S3. Example of a still snapshot of real nucleated particles in a microgravity environment.** (**A**) Real images obtained in another nucleation experiment of tungsten oxide during a parabolic flight of an aircraft of Diamond Air Service Inc., Japan. These images are shown to illustrate the existence of concentric nucleation around the entire the evaporation source, which cannot be confirmed from the images of the interference fringes because the view direction is parallel to the evaporation source. The nucleated particles are illuminated by the hot evaporation source. Because the evaporation temperatures of Fe were very high, the bright evaporation source and the darker nucleated particles could not be recorded simultaneously due to the low dynamic range of our general camera. Tungsten oxide evaporates at a much lower temperature (~1200 K) than does Fe (~2000 K), permitting observation of its nucleated particles. The number in each image shows the time in seconds. The image in the upper left (labeled as 0) corresponds to an image recorded just before the appearance of nucleated particles. (**B**) Photograph of the chamber for the microgravity experiment in the aircraft from the same viewing direction as the panel **A**. (**C**) Schematic of the evaporation source and a smoke composed of nanoparticles. The smoke forms a concentric cylinder around the filamentary evaporation source because of symmetrical diffusion of the evaporated gas under microgravity.

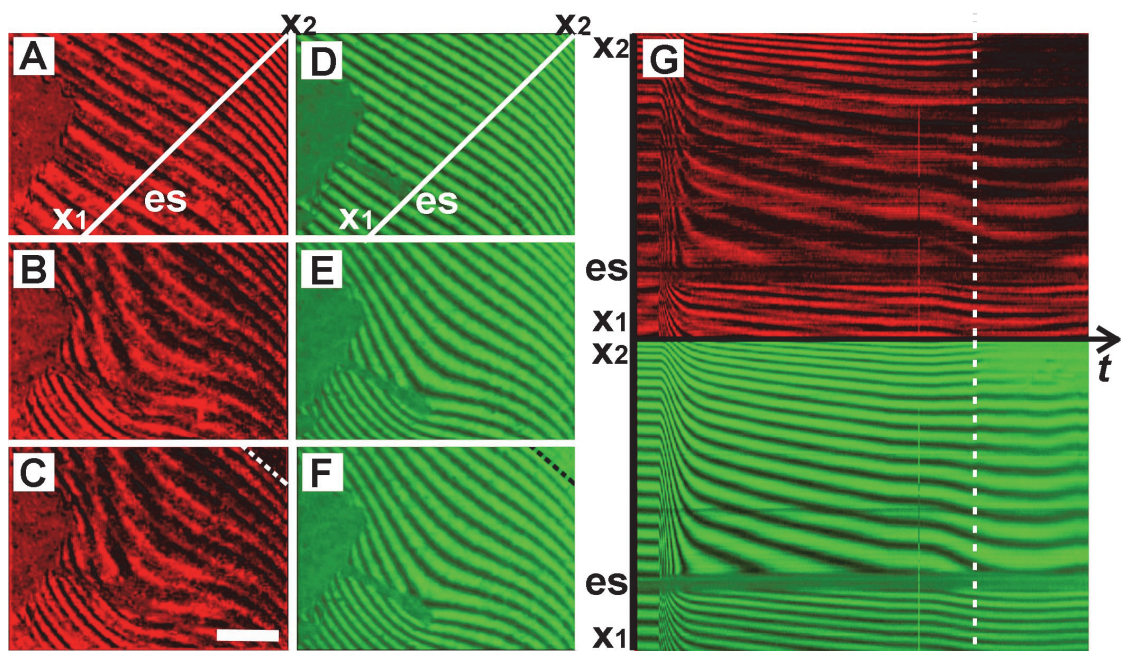

**Fig. S4. Images of interference fringes during the Fe nucleation experiment under microgravity.** Color-separated images of the interference fringes in Fig. 2 for the red laser (**A–C**) and the green laser (**D–F**) at an initial Ar buffer gas pressure of $4 \times 10^4$ Pa: (**A, D**) before heating of the evaporation source, (**B, E**) 0.4 s before the nucleation of Fe grains and (**C, F**) at the time of nucleation. In **C** and **F**, the dotted line in the upper right corner indicates the nucleation front of the Fe grains, above which interference fringes disappear due to scattering of light by abundantly formed Fe grains. Panel (**G**) shows time-series images of the interference fringes indicated by white lines in **A** and **D**. Horizontal and vertical axes in the panels correspond to the time and position from $x_1$ to $x_2$, respectively. Labels es show the position of the evaporation source. The dotted white line in **G** shows the time of nucleation. The scale bar in **C** corresponds to 3 mm.